\documentclass[prd,showpacs,showkeys,preprintnumbers,floatfix,groupedaddress,
nofootinbib,superscriptaddress,twocolumn]{revtex4-1}
\usepackage{hyperref} 
\usepackage{graphicx}
\usepackage{amssymb}
\usepackage{amsmath}
\usepackage{bbold}
\usepackage{epstopdf}
\usepackage{float}
\usepackage{caption}
\usepackage{slashed}
\usepackage{color}
\usepackage{verbatim}
\usepackage[singlelinecheck=false]{subcaption}

\DeclareGraphicsExtensions{.pdf,.png,.jpg}
\DeclareMathOperator{\tr}{tr}

\def\chpt{\raise0.4ex\hbox{$\chi$}PT}
\def\mhat{\widehat m}
\def\muhat{\widehat \mu}
\def\ephat{\hat \epsilon}

\def\EM{{\rm EM}}

\def\ThetaQCD{\Theta_{\rm QCD}}
\def\phys{{\rm phys}}

\begin{document}

\title{Phase structure with nonzero $\ThetaQCD$ and twisted mass fermions}

\author{Derek P. Horkel}
\email[e-mail: ]{dhorkel@uw.edu}
\affiliation{
 Physics Department, University of Washington, 
 Seattle, WA 98195-1560, USA \\
}
\author{Stephen R. Sharpe}
\email[e-mail: ]{srsharpe@uw.edu}
\affiliation{
 Physics Department, University of Washington, 
 Seattle, WA 98195-1560, USA \\
}
\date{\today}
\begin{abstract}
We determine the phase diagram and chiral condensate for lattice QCD
with two flavors of twisted-mass fermions in the presence
of nondegenerate up and down quarks, discretization errors
and a nonzero value of $\ThetaQCD$.
Although such a theory has a complex action and
cannot, at present, be simulated, 
the results are needed to understand how to tune to maximal twist
in the presence of electromagnetism, a topic discussed in a companion
paper.
We find that, in general,  
the only phase structure is a first-order transition of finite length.
Pion masses are nonvanishing throughout the phase plane except at the
endpoints of the first-order line.
Only for extremal values of the twist angle and $\ThetaQCD$
($\omega=0$ or $\pi/2$ and $\ThetaQCD=0$ or $\pi$) 
are there second-order transitions.
\end{abstract}

\maketitle

\section{Introduction}
\label{sec:intro}

In Refs.~\cite{Horkel:2014nha} and \cite{Horkel:2015xha} 
we determined the phase structure of two-flavor lattice QCD with Wilson and
twisted-mass fermions at nonvanishing lattice spacing
in the presence of the two physical sources of isospin breaking:
nondegenerate up and down quarks and electromagnetism.
These results are relevant for present simulations because
${\cal O}(a^2)$ discretization effects are of 
comparable size to those from isospin breaking
(Here $a$ is the lattice spacing).
Discretization effects can thus significantly
distort the chiral condensate and lead to unphysical phase transitions.

In particular, the CP-violating phase found by
Dashen in the continuum~\cite{Dashen:1970et} 
can be enlarged by discretization effects,
and for large enough $a$ can include the point with physical quark masses.

In Ref.~\cite{Horkel:2015xha} we found that
the inclusion of electromagnetism along with twisting requires one
to consider, at an intermediate stage, a lattice theory that has,
in addition to isospin breaking, a nonvanishing value of $\ThetaQCD$.
We repeat the explanation of this result below.
The purpose of the present note is to study the properties of
this extended theory, providing results that are used in 
Ref.~\cite{Horkel:2015xha} to tune to the physical value, $\ThetaQCD=0$.
We also present some results not needed in Ref.~\cite{Horkel:2015xha}
so as to provide a complete picture of the parameter dependence of
the phase structure. 

Our analysis is carried out using SU(2) chiral perturbation theory (\chpt\,).
Previous work has considered this theory at nonvanishing $\ThetaQCD$
in the continuum. In particular, Refs.~\cite{Smilga:1998dh} and
\cite{Creutz:2003xu} have studied the theory at $\ThetaQCD=\pi$,
elucidating the Dashen phase.
In addition, Refs.~\cite{Lenaghan:2001ur} and \cite{Akemann:2001ir} considered this
and related theories for arbitrary $\ThetaQCD$ in the small-volume regime,
where one can use the methods of random matrix theory. 
In all these theories they find a Dashen phase.
What our study adds to this previous work is the inclusion of the effects
of working at nonvanishing lattice spacing.

\bigskip

We begin by recalling the essential features of SU(2) \chpt\
including discretization effects, nondegenerate quarks, and electromagnetism,
in the power counting we use in 
Refs.~\cite{Horkel:2014nha} and \cite{Horkel:2015xha}.
In this power counting,
effects proportional the average light quark mass,
$m_q=(m_u\!+\!m_d)/2$, are assumed comparable to those quadratic in lattice
spacing,\footnote{%
Terms linear in $a$, if present, can, in the pion sector,
be absorbed into an additive shift in the quark mass, 
so that the leading discretization effects relevant for the
phase structure are proportional to $a^2$~\cite{Sharpe:1998xm}.}
and to those proportional to $\alpha_{\rm EM}$,
{\em i.e.}, 
$m_q \sim a^2\sim\alpha_{\rm EM}$.
We also include in the leading-order Lagrangian the dominant term introduced
by nondegeneracy, which is proportional to $\epsilon_q^2$, where
$\epsilon_q=(m_u\!-\!m_d)/2$.
We work to leading order in this combined power counting, 
so that loop effects need not be considered. 
Ignoring electromagnetism for now, the Lagrangian is then
\begin{align}
\mathcal{L}_\chi &= \frac{f^2}{4}\tr\left[
\partial_\mu \Sigma \partial_\mu \Sigma^\dagger\right]
+ {\cal V}_\chi
\\
\mathcal{V}_\chi&= -\frac{f^2}{4} 
\tr({\chi}^\dagger \Sigma + \Sigma^\dagger {\chi})
- W' [\tr(\hat A^\dagger \Sigma +   \Sigma^\dagger \hat A)]^2
\nonumber\\
&\quad +\frac{\ell_7}{16}[\tr(\chi^\dagger \Sigma - \Sigma^\dagger \chi)]^2\,.
\label{eq:Vchi}
\end{align}
Here $\Sigma\in {\rm SU(2)}$ is the chiral field, $f\approx 92\;$MeV
and $B_0$ are the continuum leading order low energy coefficients
(LECs), and $\hat A=2 W_0 a \mathbb 1$ is a spurion field, with $W_0$
and $W'$ LECs introduced by discretization errors. 
The quark mass matrix, $M$, is contained in the convenient quantity
$\chi=2 B_0 M$.\footnote{%
The detailed relationship of the masses in $M$ to the bare lattice
quark masses is explained in Appendix A of Ref.~\cite{Horkel:2015xha}.
We also note that $M$ contains only the LR projection of the full mass matrix.}
Matching physical quantities in continuum SU(2) and SU(3) \chpt\, one finds 
\begin{equation}
\ell_7 = \frac{f^2}{8 B_0 m_s}
\end{equation}
where $m_s$ is the strange quark mass~\cite{Gasser:1984gg}.
Thus we know that $\ell_7$ is positive.

In the continuum, the leading term induced in the chiral potential by electromagnetism is
that due to one-photon exchange between electromagnetic currents~\cite{Gupta:1984tb,Ecker:1988te} 
\begin{equation}
\mathcal{V}_\EM = -\frac{f^2}{4} c_\EM \tr(\Sigma \tau_3 \Sigma^\dagger \tau_3),
\label{eq:VEM}
\end{equation}
where $c_\EM$ is proportional to $\alpha_{\rm EM}$ and is known to be positive~\cite{Witten:1983ut}. 
Electromagnetism also contributes to mass renormalization, but this is implicitly included
by our use of renormalized masses in the quark mass matrix $M$. Although the
quark masses depend on the renormalization scheme chosen, this dependence
is canceled by that of the prefactor $B_0$, so that the product $\chi$ is independent
of renormalization scheme and scale.

On the lattice, with Wilson or twisted-mass fermions, the inclusion of electromagnetism
leads to additional issues. The first of these concerns the direction of the twist.
Quark nondegeneracy picks out the $\tau_3$ direction in isospin space.
In the absence of electromagnetism, one can twist  in an orthogonal
direction, e.g. $\tau_1$, and this choice leads to a real lattice
fermion determinant~\cite{Frezzotti:2003xj}.
However, such a twist leads leads to an electromagnetic current
that includes an axial component when written in terms of bare quarks.
This current cannot be coupled in a gauge-invariant way to the electromagnetic
field in a lattice theory since it is not conserved~\cite{Horkel:2015xha}.

To include both electromagnetism and nondegeneracy on the lattice, one is thus forced to
twist in the $\tau_3$ direction~\cite{deDivitiis:2011eh,deDivitiis:2013xla}.
This, however, leads to a complex lattice fermion determinant~\cite{WalkerLoud:2005bt},
making the theory challenging to simulate.\footnote{%
This is avoided in Refs.~\cite{deDivitiis:2011eh} and \cite{deDivitiis:2013xla} 
by working to linear order in perturbation theory about the
isospin symmetric theory.}
An intuitive way of understanding why the action is complex is to note that,
in the continuum, when the twist angle is $\omega$,
the fermion mass term is given by
\begin{equation}
\overline\psi (m_q c_\omega + \tau_3 \epsilon_q c_\omega
+ i\gamma_5\tau_3 m_q s_\omega +i\gamma_5 \epsilon_q s_\omega)\psi
\,,
\label{eq:tau3twist}
\end{equation}
where $\psi$ is an isodoublet,
$c_\omega=\cos\omega$, and $s_\omega=\sin\omega$.
By construction, in the continuum a nonsinglet axial rotation ({\em i.e.}, a twist)
can return the mass matrix to its standard form $m_q+\tau_3 \epsilon_q$.
However, on the lattice, such a rotation is not a symmetry.
Crudely speaking, the lattice theory with mass matrix (\ref{eq:tau3twist})
corresponds to a continuum theory in which the
coefficients of the four terms are differently renormalized.
In such a theory the mass terms involving $\gamma_5$ cannot both be rotated
away, and thus the theory has a nonzero $\ThetaQCD$.
As is well known, this leads to a complex fermion determinant.
The only redeeming feature is that, if one could tune the lattice quark mass matrix such
that it took the form of Eq.~(\ref{eq:tau3twist}) in the continuum limit, then the imaginary
part of the fermion determinant would vanish in this limit. 

The situation is not this simple, however, because of the second issue induced
by the inclusion of electromagnetism in the lattice theory. This is the presence of independent
additive renormalizations of the up and down bare untwisted quark masses proportional to
$\alpha_{\rm EM}/a$.  Since in our power counting $m\sim a^2\sim \alpha_{\rm EM}$, these
renormalizations dominate over the leading order terms described above and
collected in Eqs.~(\ref{eq:Vchi}) and (\ref{eq:VEM}).
They must be tuned away by applying nonperturbative conditions to determine,
independently, the two critical masses.\footnote{%
There is, in addition, the standard additive renormalization proportional to
$1/a$ (times powers of $\alpha_s$) that is common to both quarks. 
The nonperturbative conditions that remove the $\alpha_{\rm EM}/a$ shifts
will also remove the larger $1/a$ shifts.
The point here is that the smaller (but still divergent)
electromagnetic renormalizations imply the need for two conditions, rather than one.}
One of the results of Ref.~\cite{Horkel:2015xha} was a demonstration that the
tuning method used in Ref.~\cite{deDivitiis:2013xla} does not work in general.\footnote{%
The method, based on introducing unphysical valence quarks, works only in the
electroquenched approximation, in which sea quarks are kept neutral and degenerate.
It fails once the sea quarks are charged.
The second method proposed in Ref.~\cite{deDivitiis:2013xla}, 
and the method used in Ref.~\cite{Borsanyi:2014jba}, 
do not suffer from the same problem, because they tune using physical quantities.
}
The method provided only a single condition, while two are needed.
The key point for present purposes is that, with the untwisted parts of the quark masses ``detuned'',
the theory one is studying has, even in the continuum limit, a nonvanishing value of $\ThetaQCD$.
Thus, to come up with a second condition that will set $\ThetaQCD=0$ (in the
continuum limit) one must understand the properties of the detuned theory.
This is the purpose of the present analysis.

To understand why detuning leads to nonzero $\ThetaQCD$, it is instructive
to write out the renormalized mass matrix $M$ in a detuned, twisted theory.
It is convenient to work with $\chi$ rather than $M$,
since this is what enters the chiral Lagrangian. The form is
\begin{align}
\label{eq:chimumd}
\chi=\begin{pmatrix}
\mhat^W_u+i\muhat_u & 0\\
0 & \mhat^W_d-i\muhat_d \\
\end{pmatrix}
\end{align}
where the ``hat" on a mass indicates multiplication by $2 B_0$.
In particular,
$\mhat^W_u = 2 B_0 m^W_u$, with $m_u^W$ the renormalized
untwisted or ``Wilson'' part of the up-quark mass, while
$\muhat_u=2B_0 \mu_u$, with $\mu_u$ the twisted part of the up-quark mass.
Similar notation holds for the down-quark masses.
The superscript $W$ distinguishes the untwisted masses from the 
full physical masses, which are given, for example, by
$m_u^2 = (m_u^W)^2 + \mu_u^2$.
The two twisted masses in (\ref{eq:chimumd})
have opposite overall signs because twisting involves $\tau_3$.
In this notation, tuning the untwisted parts of both masses to their critical values 
means tuning both $\mhat^W_u$ and $\mhat^W_d$ to zero.
As long as $\mu_u$ and $\mu_d$ have opposite signs this corresponds to
tuning to maximal twist.

We can rewrite the mass matrix of Eq.~(\ref{eq:chimumd}) in
terms of the average physical quark mass $m_q$ and the nondegeneracy $\epsilon_q$:
\begin{align}
\label{eq:chiomegaphi}
\chi=\begin{pmatrix}
(\mhat_q+\ephat_q)e^{i(\varphi+\omega)} & 0\\
0 & (\mhat_q-\ephat_q)e^{i(\varphi-\omega)} \\
\end{pmatrix}\,.
\end{align}
Here $\mhat_q=2B_0 m_q$, $\ephat_q=2 B_0 \epsilon_q$,
\begin{equation}
\tan(\varphi+\omega) = \frac{\muhat_u}{\mhat_u^W}
\  \ {\rm and}\  \ 
\tan(\varphi-\omega) = - \frac{\muhat_d}{\mhat_d^W}\,.
\end{equation}
We observe from Eq.~(\ref{eq:chiomegaphi}) that $\omega$ is the twist angle,
while $\varphi$, being the overall phase of the mass matrix,
is proportional to $\ThetaQCD$:
\begin{equation}
\varphi=\frac{\ThetaQCD}2\,.
\end{equation}
Thus having a general, detuned mass matrix corresponds to working at nonzero $\ThetaQCD$.
Tuning to the critical values of the untwisted quark masses
corresponds to setting $\omega=\pi/2$ and $\varphi=0$,
{\em i.e.}, tuning to maximal twist with vanishing $\ThetaQCD$.

In summary, the dominant effect of including electromagnetism in a theory
with Wilson or twisted-mass fermions is mass renormalization.
For nonvanishing twist, this implies that one must work at nonvanishing $\ThetaQCD$ in
order to tune to $\ThetaQCD=0$. While this will be challenging for simulations,
it is straightforward to study this theory in \chpt.
Working at leading order in our power counting, one has simply to find the minima of the 
potential that is composed of the terms given in Eqs.~(\ref{eq:Vchi}) and (\ref{eq:VEM}).

\bigskip

When determining the expectation value of the chiral field, it is
convenient to parametrize it relative to the twist it would obtain
were $\varphi=a=0$:
\begin{equation}
\label{eq:sigmavev}
\langle\Sigma\rangle=
e^{i\omega \tau_3/2} e^{i\theta \hat n\cdot\vec \tau} e^{i\omega \tau_3/2}
\end{equation}
The full potential ${\cal V}={\cal V}_\chi + {\cal V}_{\rm EM}$ then becomes
\begin{align}
\label{eq:fullpot}
\begin{split}
- \frac{\mathcal{V}}{f^2} &= \mhat_q \cos{\theta}\cos{\varphi} + n_3 \ephat_q \sin{\theta}\sin{\varphi}\\
&\quad+ c_\ell \left(n_3 \ephat_q \sin{\theta}\cos{\varphi} - \mhat_q \cos{\theta}\sin{\varphi}\right)^2\\
&\quad+ w' \left(n_3 \sin{\theta} \sin{\omega} - \cos{\theta} \cos{\omega}\right)^2\\
&\quad + c_\EM \left(n_3^2 + (1-n_3^2)\cos{\theta}^2 \right)\,,
\end{split}
\end{align}
up to an irrelevant overall constant.
Here we have introduced
\begin{equation} 
c_\ell=\frac{\ell_7}{f^2}\  \  {\rm and} \  \  w'=\frac{64W'W_0^2 a^2}{f^2}.
\end{equation}
Given that $\ell_7$ and $c_{\rm EM}$ are both positive, 
the potential is always minimized with the condensate aligned in the $\tau_3$ direction, 
i.e. $\hat{n}=(0,0,\pm 1)$. 
Without loss of generality we can set $\hat{n}=(0,0,1)$ 
and absorb any sign into $\theta$.
The main task in the following is the determination of the values
of $\theta$ which minimize ${\cal V}$ as the parameters are varied.

An immediate conclusion from this analysis is that the remaining explicit
effect of electromagnetism, namely the $c_{\rm EM}$ term,
is simply a constant for $n_3=1$. It therefore does not effect the 
minimization of the potential, and thus has no impact on the phase structure.
Physically this is because the condensate lies in the neutral pion direction.
The only effect of this term is to give an overall positive shift
in the charged pion masses.\footnote{%
In light of these considerations,we drop the $c_{\rm EM}$ term in the subsequent discussion
of minimization of the potential.}

\bigskip

The remainder of this paper is organized as follows. 
In Sec.~\ref{sec:phasediagram}
we determine the phase diagram
in the $\mhat_q-\ephat_q$ plane.
We do so in stages, beginning by elucidating the
symmetries of the potential (\ref{eq:fullpot}), then working out the 
phase diagram in the continuum, next adding in discretization
effects for the extremal cases where $\omega=0$ and $\pi/2$,
and finally considering the most general choices of parameters.
We then return, in Sec.~\ref{sec:maxtwist},
to the original motivation for the present work,
namely the determination of a condition such that,
in the presence of electromagnetism, maximal twist
at $\ThetaQCD=0$ can be achieved in a physical phase.
We conclude in Sec.~\ref{sec:conclusion}.

\section{Determination of phase diagram}
\label{sec:phasediagram}

\subsection{Symmetries of the phase diagram}

Before entering into detailed calculations we
collect some general results that follow from the
form of the potential, Eq.~(\ref{eq:fullpot}).

First we note that, without loss of generality, we need only
consider $\omega$ and $\varphi$ in the range
\begin{equation}
0 \le \omega, \varphi \le \pi/2\,,
\label{eq:range}
\end{equation}
as long as we consider the full $\mhat_q-\ephat_q$ plane.
This is because ${\cal V}$ is invariant under each of the following
four transformations
\begin{align}
(i)\  & \{\omega \to\omega+\pi\} \,, \\
(ii)\  & \{\omega\to-\omega,\ \theta\to-\theta,\ \ephat_q\to-\ephat_q\}\,, \\
(iii)\ & \{\varphi\to -\varphi,\ \ephat_q\to-\ephat_q\}\,,\\
(iv)\ & \{\varphi\to \varphi+\pi,\ \mhat_q\to-\mhat_q,\ \ephat_q\to-\ephat_q\}\,.
\end{align}
In the following, we refer to the endpoints of the range (\ref{eq:range})
as the ``extremal" values of $\omega$ and $\varphi$, while values within the
range are called ``nonextremal''.

In addition, ${\cal V}$ is invariant under
\begin{align}
(v)\ & \{\omega \to \frac{\pi}2\!-\!\omega,\ \varphi \to \frac{\pi}2\!-\!\varphi,\ 
\mhat_q \leftrightarrow \ephat_q,\ \theta \to \frac{\pi}2\!-\!\theta\}\,.
\label{eq:symmv}
\end{align}
This implies relations between the phase transition lines for different values
of the parameters. For example, the phase diagram for $\omega=\varphi=0$ is
related to that for $\omega=\varphi=\pi/2$ by a reflection in the
diagonal line $\mhat_q=\ephat_q$.

The final invariance that plays a role in the following is
\begin{align}
(vi)\ & \{\omega\to \frac{\pi}2\!-\!\omega,\ \ephat_q\to -\ephat_q,\
w' \to -w',\ \theta\to -\theta\}\,.
\label{eq:symmvi}
\end{align}
This relates the phase diagrams with opposite signs of $w'$.

\subsection{Continuum \chpt\ with nonzero $\ThetaQCD$}
 
In this section we examine the phase structure in the continuum.
Without discretization effects, the twist angle 
is redundant and has no effect on the phase diagram. 
This is manifest in the basis used in Eq.~(\ref{eq:fullpot}), 
where with $w' \propto a^2=0$ there is no dependence on $\omega$.

We begin by recalling results for the extremal cases $\varphi=0$ and $\pi/2$, corresponding to $\ThetaQCD=0$ and $\pi$. These have real and imaginary quark masses, respectively.

\begin{figure*}
\centering
 \begin{subfigure}{0.49\textwidth}
\includegraphics[scale=.35]{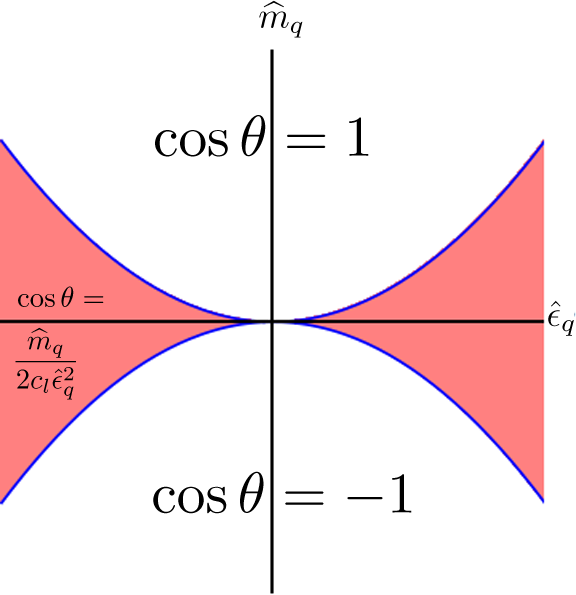}
\caption{\label{fig:phi0}$\varphi = 0$}
\end{subfigure}
 \begin{subfigure}{0.49\textwidth}
\includegraphics[scale=.35]{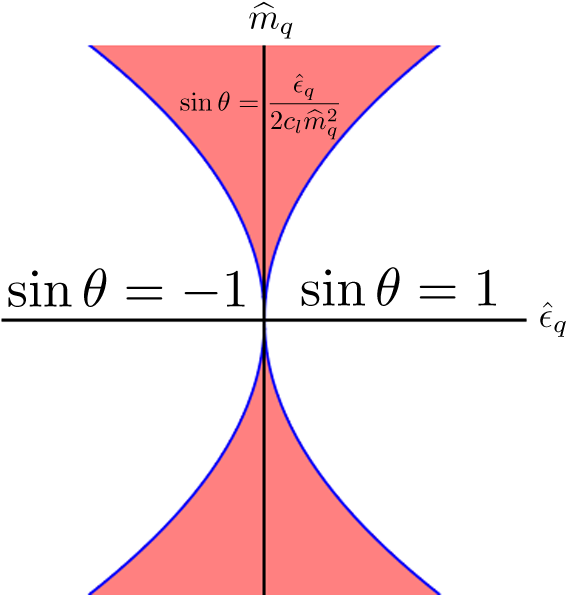}
\caption{\label{fig:phimax} $\varphi = \pi/2$}
\end{subfigure}
\caption{ 
\label{fig:cont}
Continuum phase diagram for (a) $\varphi = \ThetaQCD/2= 0$ and (b) $\varphi= \pi/2$. 
Shaded (pink) regions have varying values of the vacuum angle $\theta$, as indicated in the figures.
Unshaded regions have constant $\theta$. 
The neutral pion mass vanishes along the phase transition lines.}
\label{fig:fig1}
\end{figure*}

The physical case, $\varphi=0$, has been described extensively in the 
literature~\cite{Dashen:1970et, Creutz:2011hy, Horkel:2014nha}. 
The phase diagram is shown in Fig.~\ref{fig:fig1}(a). 
There is a second-order transition between the standard continuum phase
and the CP-violating Dashen phase, lying along $\mhat_q=\pm2c_\ell \ephat_q^2$. 
In the Dashen phase, the potential has two degenerate minima, both having
\begin{equation}
\cos{\theta}=\frac{\mhat_q}{2 c_\ell \ephat_q}\,,
\end{equation}
and differing in the sign of $\theta$.

The case of $\ThetaQCD=\pi$ ($\varphi=\pi/2$)
was first described by Smilga~\cite{Smilga:1998dh}.
The potential has the same form as for $\varphi=0$, 
except that $\mhat_q$ and $\cos{\theta}$ are exchanged with $\ephat_q$ and $\sin{\theta}$, respectively. 
This implies that the phase diagram has the same form as for $\varphi=0$, except that it
is reflected in the $\mhat_q=\ephat_q$ line, as shown in Fig.~\ref{fig:fig1}(b).
This is an example of the symmetry (\ref{eq:symmv}) at work (since the change
in $\omega$ is irrelevant in the continuum).
There is thus a second-order transition to a Dashen-like phase along the lines 
$\ephat_q=\pm2c_\ell \mhat_q^2$. There is again a two-fold degeneracy within this phase.

For nonextremal $\varphi$ ($0<\varphi < \pi/2$) the potential 
(\ref{eq:fullpot}) cannot be minimized exactly and it is instructive to look at some simple limits.

First we drop the $\mathcal{O}(m^2)$ $c_\ell$ term.
In the extremal cases this means that the width of the shaded (pink) phases shrinks to zero,
so that there is a first-order transition along the entire
$\mhat_q=0$ line for $\varphi=0$ and 
along the $\ephat_q=0$ line for $\varphi=\pi/2$.
By contrast, for nonextremal $\varphi$, there are no transitions.
The potential is minimized at,
\begin{equation}
\tan{\theta}=\frac{\ephat_q}{\mhat_q}\tan{\varphi}\,,
\label{eq:phiLO}
\end{equation}
and changes continuously as one moves through the phase diagram,
except when passing through the origin.

The absence of a transition for nonextremal $\varphi$ continues to hold when the $c_\ell$ term is
restored. This can be understood as due to the lack of a $Z_2$ symmetry in the potential. 
It is the presence of a $Z_2$ symmetry for extremal $\varphi$ 
(under which $\theta\to \theta+\pi$) that, 
when broken by the vacuum, leads to a second-order transition.
The upshot is that the extremal phase diagrams of Fig.~\ref{fig:fig1} 
are replaced by blank diagrams
with no transitions, aside from the singular point at the origin.

To show a concrete example of this, we consider $\varphi=\pi/4$. 
Using the parametrization $\mhat_q=r \cos{\alpha}$, 
$\ephat_q=r \sin{\alpha}$ and $r^2 =\mhat_q^2+\ephat_q^2$. 
The potential is then
\begin{equation}
- \frac{\mathcal{V}}{f^2} = \frac{r}{\sqrt{2}} 
\left[\cos{(\theta-\alpha)} + \kappa \cos^2{(\theta+\alpha)}\right]\,,
\end{equation}
where $\kappa={c_\ell r}/{\sqrt{2}}$ can be treated as small in our power counting. 
The minima occur when
\begin{equation}
0 = \sin{(\theta-\alpha)} + 2\kappa \sin{[2(\theta+\alpha)]}\,.
\end{equation}
Expanding in powers of $\kappa$ about the leading-order solution, $\theta=\alpha$, 
we find
\begin{equation}
\theta=\alpha-2\kappa\sin{(4\alpha)} +\mathcal{O}(\kappa^2)\,.
\end{equation}
The presence of only a single solution indicates the absence of a Dashen-like phase.
We have investigated this numerically for other values of $\varphi$ and 
found that there are no phase transitions for any nonextremal $\varphi$.

To see how the degeneracy of the Dashen-like phase is broken for nonextremal $\varphi$, 
consider the potential along the $\mhat_q=0$ axis:
\begin{equation}
- \frac{\mathcal{V}}{f^2} = \ephat_q \sin{\theta}\sin{\varphi} + 
c_\ell \ephat_q^2 \sin^2{\theta}\cos^2{\varphi}\,.
\end{equation}
For $\varphi=0$, one finds (since $c_\ell>0$) that there are degenerate minima at $\sin{\theta}=\pm 1$.
This corresponds to moving from the origin in Fig.~\ref{fig:fig1}(a) along the $\ephat_q$ axis and
thus lying in the (shaded pink) Dashen phase.
Turning on a nonzero $\varphi$, the potential is still extremized at 
$\left\vert\sin\theta\right\vert=1$, but the two
extrema are no longer degenerate
\begin{equation}
- \frac{\mathcal{V}(\sin\theta=\pm 1)}{f^2} = \pm \ephat_q \sin{\varphi} + c_\ell \ephat_q^2\cos^2\varphi \,.
\end{equation}
Thus there is a unique minimum, such that $\sin\theta=1$ for $\ephat_q>0$ and $\sin\theta=-1$ 
for $\ephat_q<0$ (assuming a positive $\varphi$). There thus can be no Dashen-like phase.

\subsection{Discretization effects at nonzero $\ThetaQCD$ for extremal $\omega$}

\begin{figure*}
\centering
\begin{subfigure}{0.49\textwidth}
\includegraphics[scale=.35]{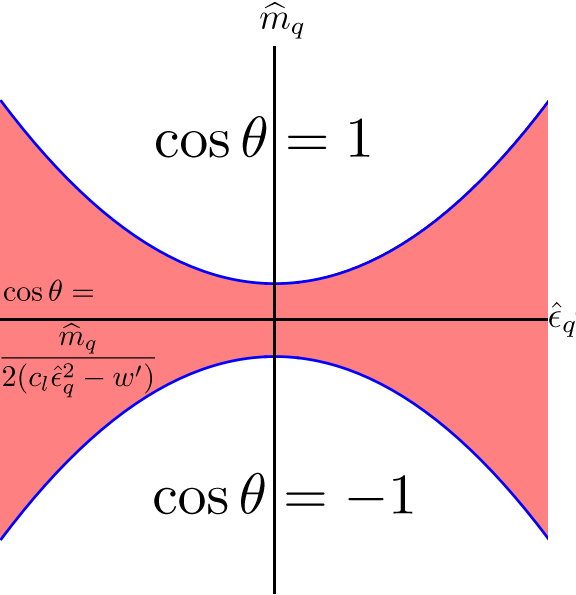}
\caption{\label{fig:Aokiold} $w'<0$.}
\end{subfigure}
\begin{subfigure}{0.49\textwidth}
\includegraphics[scale=.35]{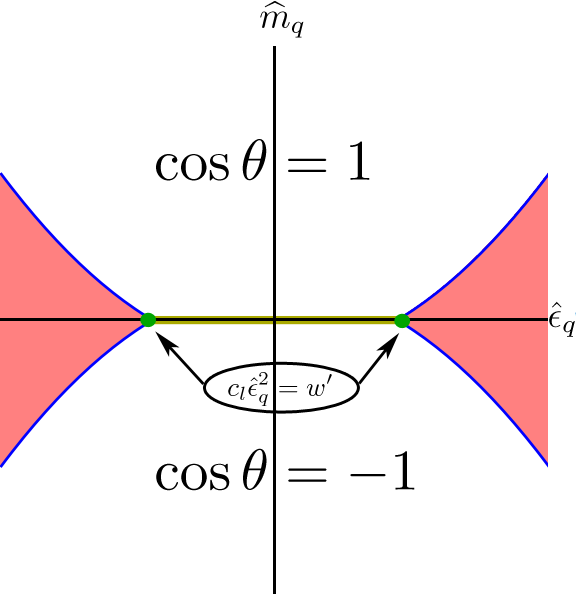}
\caption{\label{fig:Firstold} $w'>0$.}
\end{subfigure}
\caption{\centering Phase diagrams from Ref.~\cite{Horkel:2014nha}
including effects of discretization for $\omega=\varphi=0$: (a) Aoki scenario ($w'<0$) and
(b) first-order scenario ($w'>0$)..
The expression for $\theta$ in the shaded (pink) region in (a) also holds in (b).
The boundary of the shaded regions are second-order transition lines,
along which the neutral pion mass vanishes. 
The (yellow) solid line running along the $\ephat_q$ axis between the
shaded regions is a first-order transition.}
\label{fig:fig2}
\end{figure*}

We now turn on discretization errors by considering non vanishing $w'$. 
Just as in the continuum, the phase diagram is easiest to determine for extremal $\varphi$. 
The case of $\omega=\varphi=0$ (untwisted fermions with $\ThetaQCD=0$) has long been studied, 
and it has been shown that there are two distinct scenarios depending on the sign of $w'$: 
the so-called Aoki scenario for $w'<0$, and the first-order scenario for
$w'>0$~\cite{Aoki:1983qi,Sharpe:1998xm,Horkel:2014nha}. 
The resulting phase diagrams are shown in Fig.~\ref{fig:fig2}, and should be compared to the continuum diagram of Fig.~\ref{fig:fig1}(a).
For $w'<0$,  the Dashen phase, in which $\theta$ is degenerate, expands vertically
so as to include the origin. The CP violating phase along the $\mhat_q$ axis is typically called the Aoki phase, so we call the extended CP violating region the Aoki-Dashen phase. This situation is shown in Fig.~\ref{fig:fig2}(a).
For $w'>0$, the vertical width of the continuum Dashen phase is reduced,
and there is a segment of first-order transition along the $\ephat_q$ axis,
as shown in Fig.~\ref{fig:fig2}(b). In both scenarios,
within the Aoki-Dashen phases the potential is minimized by
\begin{equation}
\cos{\theta}=\frac{\mhat_q}{2 (c_\ell \ephat_q -w')}\,,
\end{equation}
so that again there are two degenerate vacua with opposite signs of $\theta$.

\begin{figure*}
\centering
\includegraphics[scale=.5]{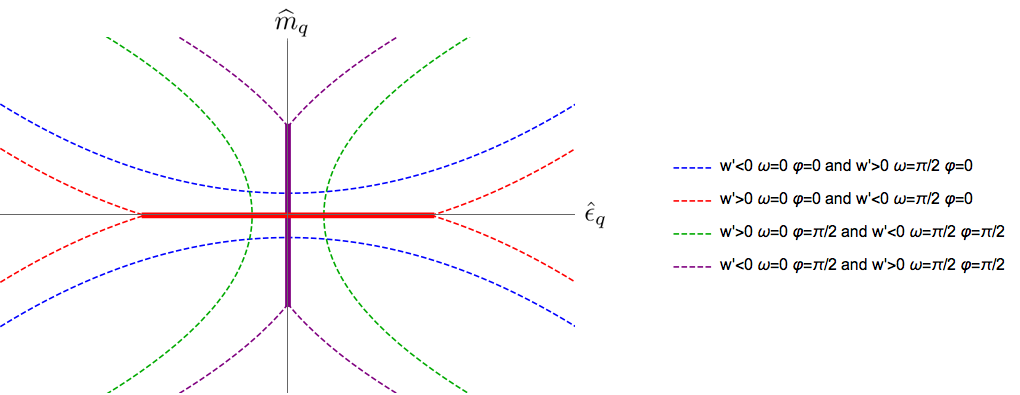}
\caption{\centering 
Schematic positions of phase boundaries for extremal choices of $\omega$ and $\varphi$ 
for both $w'<0$ and $w'>0$ scenarios.
Dashed lines indicate second order transitions, solid lines indicate first-order transitions. 
Results for two of these parameter choices also appear in Fig.~\ref{fig:fig2}.}
\label{fig:fig3}
\end{figure*}

We next consider $\varphi=\pi/2$ while holding $\omega=0$ (i.e. Wilson fermions at $\ThetaQCD=\pi$).
This has not been previously studied in the presence of lattice
artifacts.
As described above, for the continuum terms in ${\cal V}$, changing $\varphi$ from $0$ to $\pi/2$ has
the effect of interchanging $\mhat_q$ and $\cos{\theta}$ with $\ephat_q$ and $\sin{\theta}$, respectively.
Since the $w'$ term can be rewritten as $w'\cos^2\theta=w'(1-\sin^2\theta)$,
the same interchanges hold for $w'\ne 0$ as long as one flips the sign of $w'$.
Up to some unimportant sign flips,
this is an example of the general transformation 
obtained by combining Eqs.~(\ref{eq:symmv})
and (\ref{eq:symmvi}):
\begin{align}
(vii)\ & \{\varphi\to\frac{\pi}2\!-\! \varphi,\
\mhat_q\to -\ephat_q,\
\ephat_q\to \mhat_q,
\nonumber \\ 
&\qquad\qquad w'\to-w',\
\theta\to\theta-\frac{\pi}2\}\,.
\label{eq:symmvii}
\end{align}
The implication is that the phase diagrams for $\varphi=\pi/2$ 
are obtained from those of Fig.~\ref{fig:fig2} by rotating 9$0^\circ$
counterclockwise,
and interchanging the $w'<0$ and $w'>0$ scenarios.
The positions of the resulting
transitions are shown schematically in Fig.~\ref{fig:fig3}.

The twist angle is no longer redundant when $w'\ne 0$,
entering the $w'$ term  in Eq.~(\ref{eq:fullpot}) as $w' \cos^2(\theta+\omega)$.
Thus changing $\omega$ from $0$ to $\pi/2$ 
has the effect of flipping the sign of $w'$:
$w'\cos^2(\theta+\pi/2)=w'\sin^2{\theta}=w'(1-\cos^2{\theta})$. 
This is an example of the general transformation (\ref{eq:symmvi}).
It implies that the phase diagrams for maximal twist 
can be simply obtained from those without twist.
The situation is summarized in Fig.~\ref{fig:fig3}.

\begin{figure*}
\includegraphics[scale=.6]{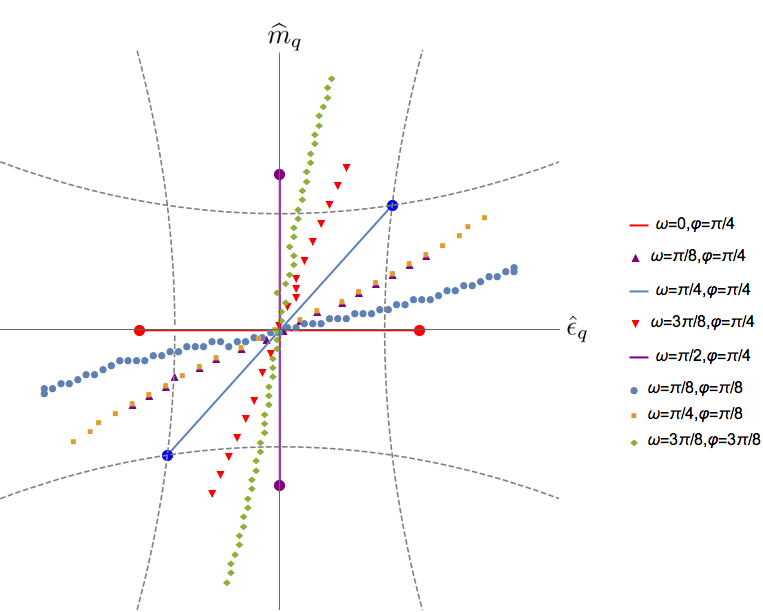}
%
\caption{\label{fig:fig4} 
Phase diagram including discretization effects (with $w'>0$)
for several values of $\omega$ and $\varphi$. 
Solid lines are analytically determined first-order 
transition lines described in the text. 
Points represent the location of the numerically 
determined first-order transition lines.
Dashed lines show the positions of the central second-order transition lines that arise at extremal values of
$\omega$ and $\varphi$ as shown in Fig.~\ref{fig:fig3}.
These are included to set the scale, since they depend on the values of $w'$ and $c_\ell$.
Results for $w'<0$ can be obtained from these using the
transformations of Eqs.~(\protect\ref{eq:symmvi}) and 
(\protect\ref{eq:symmvii}).}
\end{figure*}

In the remainder of this subsection we keep $\omega$ at 
one of the extremal values but allow $\varphi$ to take on nonextremal values. 
We recall that in the continuum, the phase
diagram with such parameters has no phase transitions.
This turns out not to be the case when $w'\ne 0$.
Examples of the results we find are shown in Fig.~\ref{fig:fig4}.

We begin with $\omega=0$ and nonextremal $\varphi$,
and work in the $w'>0$ scenario.
We find that there is a first-order transition along a finite segment of the 
$\ephat_q$ axis, across which $\theta$ changes discontinuously.
The length of the segment depends on $\varphi$.
As $\varphi$ approaches zero [in which limit one obtains the phase diagram of Fig.~\ref{fig:fig2}(b)]
the first-order segment asymptotes to precisely the 
first order transition line shown in Fig.~\ref{fig:fig2}(b),
with end points $c_\ell \ephat_q^2 = \pm w'$.
We stress again that, for non vanishing $\varphi$, there are no regions
of Aoki-Dashen phase.
As $\varphi$ increases, the first-order segment reduces in length,
until, as $\varphi\to\pi/2$, it approaches the width of the Aoki phase that
appears at $\varphi=\pi/2$, i.e. with end points $\ephat_q=\pm 2 w'$.
[Recall that the phase diagram at $\varphi=\pi/2$ is given by
Fig.~\ref{fig:fig2}(a) rotated by $90^\circ$; see also Fig.~\ref{fig:fig3}.]
The first-order segment at the halfway point, $\varphi=\pi/4$,
is shown as the horizontal solid (red) line in Fig.~\ref{fig:fig4}.

The length of the segment can be obtained analytically for all $\varphi$.
To do so, one extremizes the potential after setting $\mhat_q=0$. 
The global minimum lies at
\begin{equation}
\sin{\theta}=\frac{-\ephat_q \sin{\varphi}}{2(c_\ell \ephat_q^2 \cos^2{\varphi}-w')}
\,,
\end{equation}
with the sign of $\cos\theta$ undetermined.
As one passes through the transition line (by varying $\mhat_q$),
$\cos\theta$ changes sign, indicating a first-order transition.
Solving for the endpoints, where $\cos{\theta}=\pm 1$, we find
\begin{equation}
\left| \ephat_q \right| = \frac{-\sin{\varphi}+
\sqrt{\sin^2{\varphi}+16c_\ell w' \cos^2{\varphi} }}{4c_\ell \cos^2{\varphi}} \,.
\end{equation}
This gives the results quoted above in the limits $\varphi\to 0, \pi/2$.

The corresponding results for $w'<0$ can be obtained from those just described for $w'>0$ using the transformation of Eq.~(\ref{eq:symmvii}).
In words, to obtain the phase diagram for $\varphi=\varphi_0$ and $w'=w'_0<0$,
one takes the diagram with $\varphi=\pi/2-\varphi_0$ and $w'=|w'_0|$ and
rotates it by $90^\circ$ counterclockwise. This implies that the first-order transition
line is now vertical.

Similarly, one can obtain results for $\omega=\pi/2$ from those at $\omega=0$
using the transformation of Eq.~(\ref{eq:symmv}).
Specifically to obtain the phase diagram for $\varphi=\varphi_0$ at $\omega=\pi/2$,
one takes that with $\varphi=\pi/2-\varphi_0$ and $\omega=0$ and reflects it
in the $\mhat_q=\ephat_q$ line.
This implies that the first order line is vertical for $w'>0$ and horizontal
for $w'<0$. An example of this result (for $w'>0$) is shown by the
vertical solid (purple) line in Fig.~\ref{fig:fig4}.

Since there are no second-order phase transitions, the pion masses are
nonvanishing throughout the phase plane, with the exception of the
endpoint of the first-order transitions, where the mass of the neutral pion vanishes.

%

\subsection{Nonextremal $\omega$ and $\varphi$}
\label{sec:numb}

Finally, we consider the most general choice of parameters: both
$\omega$ and $\varphi$ nonextremal. Here, in most cases, we have
to proceed numerically, although we can check
the results analytically for the special case of $\omega=\varphi=\pi/4$.

We have found in the previous subsection that, for extremal $\omega$
but nonextremal $\varphi$ there is a first-order transition line of finite
length that is oriented either horizontally or vertically. For example,
Fig.~\ref{fig:fig4} shows that for $\varphi=\pi/4$ and $w'>0$, the
transition line is horizontal at $\omega=0$ and vertical at
$\omega=\pi/2$.
It is not surprising, therefore, that for intermediate values of $\omega$
there is a first order transition line of finite length at an intermediate
angle that interpolates between the horizontal and vertical limits. 
Examples for several intermediate values of $\omega$ for
$\varphi=\pi/4$, $\pi/8$ and $3\pi/8$ are shown in the figure.
We observe that, aside from the special case of $\omega=\varphi=\pi/4$,
the first order lines are ``S-shaped'' rather than straight.
We also observe an example of an overlapping transition line
(though of different lengths) for the parameter choices
$(\omega,\varphi)=(\pi/8,\pi/4)$ and $(\pi/4,\pi/8)$. We have not
understood this overlapping analytically, and do not know if it is exact.
Figure~\ref{fig:fig4} also shows an example of the application of
the symmetry of Eq.~(\ref{eq:symmv}), which implies that
the transition lines for parameters
$(\omega,\varphi)=(\pi/8,\pi/8)$ and $(3\pi/8,3\pi/8)$ should be
related by reflection in the diagonal $\mhat_q=\ephat_q$ line. 

For $\omega=\varphi = \pi/4$, we know from the symmetry of
Eq.~(\ref{eq:symmv}) that the transition line must be invariant under
reflection in the $\mhat_q=\ephat_q$ line.
Thus it must either lie along this line or be perpendicular to it, and
in both cases it must be straight.
It turns out that, for $w'>0$ it lies along the diagonal, as shown
in Fig.~\ref{fig:fig4} by the solid (blue) line.
Given this information, it is straightforward to determine the
end points analytically, and we find that they lie at
\begin{equation}
\mhat_q=\ephat_q=\pm \frac{1-\sqrt{1-16 c_\ell w'}}{4 c_l}\,.
\end{equation}
We note the curious result
that these points lie at the junction of the
boundaries of the Aoki-Dashen phases for extremal $\omega$ and $\varphi$.

\section{Maximal twist condition}
\label{sec:maxtwist}

The standard technique for tuning to maximal twist in the absence of 
isospin breaking is to enforce the vanishing of the ``PCAC mass".
This determines the critical value of the untwisted component of the common quark mass.
This critical value can then be used even in the presence of isospin
breaking due to quark masses, i.e. when the twisted components of the
up and down quark masses differ. 
This is no longer the case when electromagnetism is included, because,
as explained in the Introduction,
the critical masses for the up and down quarks differ.
Setting the PCAC mass to zero is essentially a way of enforcing, in a particular
correlation function, the restoration of SU(2) flavor and parity symmetries at nonzero
lattice spacing. In the presence of electromagnetism, however,
these symmetries are absent even in the continuum limit, so it makes no sense to enforce them.
Thus one must use alternative methods to tune to maximal twist.

In our companion paper, Ref.~\cite{Horkel:2015xha}, 
we analyze a method for carrying out the tuning in the presence of electromagnetism,
proposed in Ref.~\cite{deDivitiis:2013xla},
This involves partial quenching, and our analysis is somewhat involved, but the
details do not matter here. Our key finding is that the method fails to
tune the untwisted components of the up and down quark masses to zero,
as required for maximal twist, but rather only enforces 
a condition on the condensate:
\begin{equation}
\langle \Sigma \rangle\equiv e^{i(\theta+\omega)\tau_3}= e^{i (\pi/2) \tau_3}\  \ \Rightarrow\ \
\theta'\equiv \theta+\omega = \pi/2\,.
\label{eq:condition}
\end{equation}
In the first equality we are simply using the definition of the phase angle $\theta$,
given in Eq.~(\ref{eq:sigmavev}), together with the result that $\vec n$ points in the $\tau_3$
direction. This implies that the total $\tau_3$ rotation angle is $\theta'=\theta+\omega$, which
is set by the condition of Ref.~\cite{deDivitiis:2013xla} to $\pi/2$.

The condition (\ref{eq:condition}) is indeed consistent with the desired parameters, 
i.e. with $\omega=\pi/2$ and $\varphi=0$. To see this we note that, for these parameters,
the phase diagrams are those of
Fig.~\ref{fig:fig2} except that the $w'<0$ and $w'>0$ diagrams are interchanged.\footnote{%
This  result is obtained by acting with the transformation (\ref{eq:symmvi}) on the $\omega=\varphi=0$
results that are actually shown in the figure. See also also Fig.~\protect\ref{fig:fig3}.}
Thus, as long as the physical masses are such that one is in the unshaded region,
i.e. as long as one avoids the Aoki-Dashen phases, one has that $\cos\theta=1$ and
thus $\theta=0$. This means that $\theta'=\omega=\pi/2$, which satisfies Eq.~(\ref{eq:condition}).

However, there is in general a one-dimensional family of solutions to Eq.~(\ref{eq:condition}),
all having different values of $\omega$ and $\varphi$. One can understand this intuitively
as follows. Nonvanishing values of $\omega$ and of $\varphi$ both violate parity, and thus
both lead to a nonzero twist of the condensate, i.e. a nonzero value of $\theta'$.
For any choice of $\varphi$,
the desired value $\theta'=\pi/2$ can, in general,
be obtained by a suitable value of $\omega$.
Thus there is a line in the $\omega-\varphi$ plane along which the condition is satisfied.
In order to tune to the desired point on this line an additional condition is needed.

It turns out to be easier to do the calculation using the parametrization 
of the mass matrix given in Eq.~(\ref{eq:chimumd}).
Here we fix the twisted components of the masses $\muhat_u$ and $\muhat_d$ 
(ultimately to their physical values,
namely $2 B_0 m_u^\phys$ and $2 B_0 m_d^\phys$, respectively) and vary
the untwisted ccmponents
$\mhat_u^W$ and $\mhat_d^W$. This corresponds to what is done in actual simulations.
The condition of Eq.~(\ref{eq:condition}) then forces the
theory to lie on a line in the $\mhat_u^W-\mhat_d^W$ plane. Our aim is to determine
this line and to find an additional condition that picks out the desired point on the line,
namely $\mhat_u^W=\mhat_d^W=0$.

\begin{figure}[h]
\centering
\includegraphics[scale=.35]{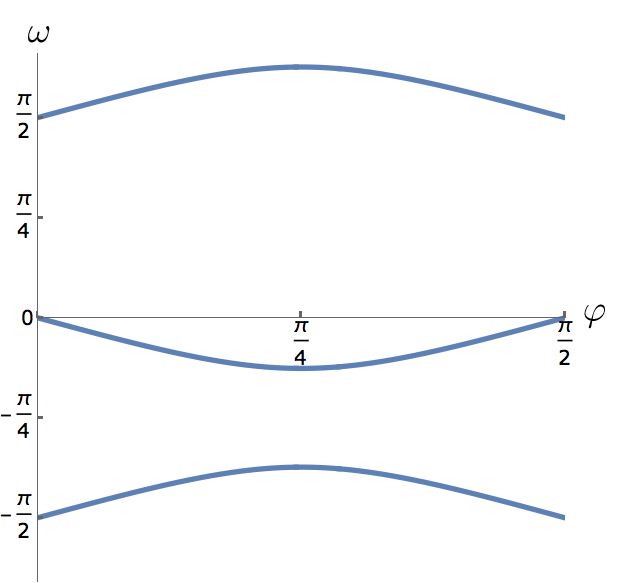}
\caption{ 
\label{fig:fig5}
Values of $\omega$ and $\varphi$ for which $\theta'=\pi/2$, using 
$\muhat_u\approx 2 B_0 m_u^\phys$ and  $\muhat_d\approx 2 B_0 m_d^\phys$,
but with $c_\ell$ larger than the physical value so as to increase
the curvature of the lines for the sake of clarity.}
\end{figure}

In terms of the parametrization (\ref{eq:chimumd}) the potential is
\begin{align}
\label{eq:mumdpot}
- \frac{\mathcal{V}}{f^2} &= \frac{\mhat_u^W+\mhat_d^W}{2} \cos{\theta'} 
+  \frac{\muhat_u+\muhat_d}{2}\sin{\theta'}\nonumber\\
&\quad+ c_\ell \left(\frac{\muhat_u-\muhat_d}{2}\cos{\theta'}
- \frac{\mhat_u^W-\mhat_d^W}{2}\sin{\theta'}\right)^2\nonumber\\
&\quad+ w' \cos^2{\theta'}\,.
\end{align}
In order for an extremum of this potential to lie at $\theta'=\pi/2$, 
it is simple to show that the untwisted masses must satisfy
\begin{equation}
\label{eq:sdef}
\frac{\mhat_d^W}{\mhat_u^W} 
= - \left( \frac{1-c_\ell (\muhat_u-\muhat_d)}{1+c_\ell (\muhat_u-\muhat_d)} \right)
\equiv s \,,
\end{equation}
{\em i.e.}, the theory must lie along a straight line in the 
$\mhat_u^W-\mhat_d^W$ plane with slope $s$
determined by the physical masses and $c_\ell$.
We can turn this into a constraint on $\omega$ and $\varphi$ by equating the
parametrizations of Eqs.~(\ref{eq:chiomegaphi}) and~(\ref{eq:chimumd}).
One finds $\mhat_{u,d}^W=\muhat_{u,d} \cot{(\omega \pm \varphi)}$,
so that the allowed values of $\omega$ and $\varphi$ satisfy
\begin{equation}
s \; \muhat_u \cot{(\omega+\varphi)} = \muhat_d \cot{(\omega-\varphi)}\,.
\end{equation}
Lines satisfying this equality 
are shown in Fig.~\ref{fig:fig5}.
The desired point is at $\varphi=0$, $\omega=\pi/2$, but, as claimed above,
solutions exist for all values of $\varphi$.

The above considerations assume that the extrema at $\theta'=\pi/2$ is a minimum
of the potential. This can be determined by examining either the second derivative of the potential evaluated at the correct value of $s$ and $\theta'=\pi/2$ or, equivalently, 
by checking that the neutral pion mass is nonnegative. 
The neutral pion mass along the line (\ref{eq:sdef}) is
\begin{align}
m_{\pi^0}^2=& \frac{\hat \mu_u\!+\!\hat \mu_d}2
\!-\!2 c_\ell \left(\frac{\hat \mu_u \!-\! \hat \mu_d}2\right)^2
\!\!\!+\! 2 c_\ell \left(\frac{\hat m_u^W\!-\! \hat m_d^W}2\right)^2
\!\!\!-\!2 w'.
\label{eq:mpi0constrained}
\end{align}
The sum of the first two terms is positive for physical parameters (since this is just
the physical neutral pion mass-squared at this order in \chpt, and higher order corrections are small).
The third term is always positive and vanishes only at the point of maximal twist
(as long as $s\ne 1$, which is the case for physical parameters).
The last term can be negative if $w'>0$. Thus, if $w'$ takes a large enough positive value,
it can be that the point we are aiming to tune to does not lie at the minimum of the potential.
This happens when the physical point lies inside the Aoki-Dashen phase.

Assuming that this does not happen, we can ask what criterion can be used to
tune to maximal twist along the lines satisfying Eq.~(\ref{eq:condition}). The criterion
proposed in Ref.~\cite{Horkel:2015xha}
is simply to minimize the neutral pion mass, Eq.~(\ref{eq:mpi0constrained}), since, as already
noted, this occurs when $\hat m_u^W=\hat m_d^W=0$. One can also minimize the charged
pion mass, the expression for which is given in Ref.~\cite{Horkel:2015xha}.

\bigskip
We close this section by making a connection with our results for the phase diagram
for general $\omega$ and $\varphi$, obtained in Sec.~\ref{sec:phasediagram}.
In particular, we imagine that we have somehow tuned close to maximal twist, but that
there is a small offset. Specifically we fix 
$\omega=\pi/2+\delta$ and $\varphi=\alpha \delta$ with $|\delta|\ll 1$ and
$\alpha\sim {\cal O}(1)$. 
This differs from (and is less realistic than) our analysis above where we fixed the
twisted up and down masses.
Nevertheless, this allows a valid theoretical exercise: with $\omega$ and $\varphi$ fixed in
this way, we determine the line in the
$\mhat_q-\ephat_q$ plane that satisfies the tuning condition of Eq.~(\ref{eq:condition}).

\begin{figure*}
\centering
 \begin{subfigure}{0.49\textwidth}
\includegraphics[scale=.4]{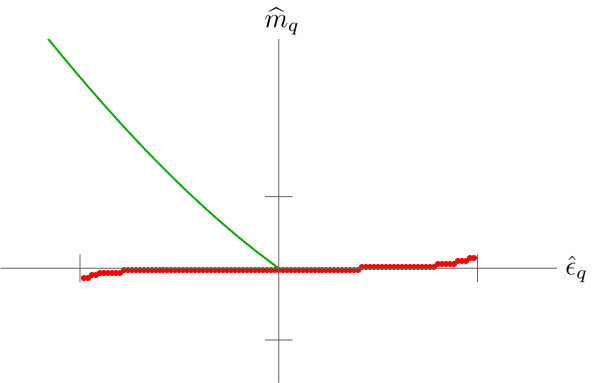}
\caption{\label{fig:ExampleAoki}}
\end{subfigure}
 \begin{subfigure}{0.49\textwidth}
\includegraphics[scale=.4]{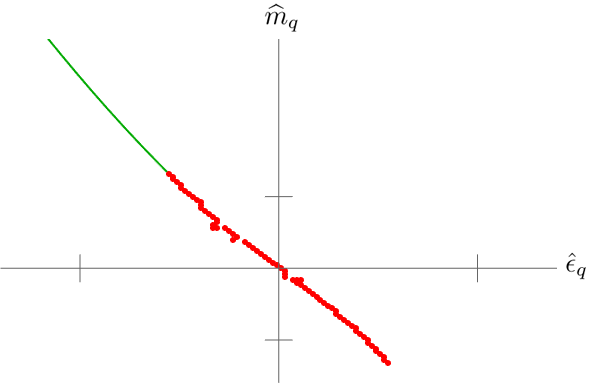}
\caption{\label{fig:ExampleFirst}}
\end{subfigure}
\caption{\label{fig:fig6} 
Applying the tuning condition at fixed $\omega=\pi/2+0.1$ and $\varphi=0.05$
for (a) $w'<0$ and (b) $w'>0$.
The size of $c_\ell$ and $w'$ can be seen from the tick marks on the axes,
which lie at $\mhat_q=\pm 2w'$ and $\ephat_q=\pm \sqrt{\left|w'\right|/c_\ell}$.
The solid (green) line shows the result of applying the condition (\protect\ref{eq:condition})
as well as requiring that $m_{\pi^0}^2\ge 0$.
(Red) points show the locations of the numerically determined first-order transition lines.}
\end{figure*}

Written in terms of the variables of Eq.~(\ref{eq:chiomegaphi}), the
tuning condition becomes
\begin{equation}
s (\mhat_q+\ephat_q) \cos(\omega+\varphi) = (\mhat_q-\ephat_q) \cos(\omega-\varphi)\,,
\end{equation}
where we recall that $s$ is given by Eq.~(\ref{eq:sdef}).
For our fixed values of $\omega$ and $\varphi$, this equation can be converted into
a result in the $\mhat_q-\ephat_q$ plane, using an expansion in powers of $\delta$:
\begin{equation}
\label{eq:maxtwistmq}
\mhat_q = -\ephat_q \frac{\alpha-2c_\ell \ephat_q}{1-2\alpha c_\ell \ephat_q} + \mathcal{O}(\delta^2)\,.
\end{equation}
Examples of this result (for both signs of $w'$) 
are shown in Fig.~\ref{fig:fig6} by the (green) solid lines.
As noted above, this result is only valid if the pion mass-squared of
Eq.~(\ref{eq:mpi0constrained}) is positive or zero. Thus the line terminates at the point where
$m_{\pi^0}$ vanishes, which occurs only for $w'>0$.

Also shown in the figures are the positions of the first-order lines, which have been determined
numerically. We observe that, for $w'<0$, the line along which Eq.~(\ref{eq:condition}) holds
goes all the way to the origin, where it runs into the first-order line.
By contrast, for $w'>0$ the endpoint of the tuned line is precisely the starting point of
the first-order line. This is reasonable since it is the only position in the phase diagram
where a pion is massless. In any case, we see that, even for nonextremal $\omega$ and
$\varphi$, where there is only a first-order transition, 
the condition (\ref{eq:condition}) cannot be maintained
all the way to the origin.

\section{Conclusions}
\label{sec:conclusion}

In this short note, we have determined the phase structure of lattice QCD in
the presence of isospin breaking and a nonvanishing value of $\ThetaQCD$.
This is, for the present, a theoretical exercise, but one that was necessary 
in order to understand how to tune to maximal twist in the presence of electromagnetism,
an analysis that was completed in our companion paper~\cite{Horkel:2015xha}.

The results are also interesting in their own right. In particular, for generic 
(nonextremal) values of the twist angle and $\ThetaQCD$, the continuum theory has no
phase structure, while the lattice theory has a segment of first-order transition whose length
is set by $w'$ and is thus of ${\cal O}(a^2)$. 

We have kept in our analysis only the leading order terms arising from each type of symmetry
breaking. A quantitative analysis would require the inclusion of all other terms of 
$\mathcal{O}(m^2)$ as well as those proportional to $m a$. 
Based on our work in Ref.~\cite{Horkel:2014nha}, however, we do not expect
these terms to lead to qualitative changes in the phase diagrams.

\section{Acknowledgements}
\label{sec:acknowledgements}

This work was supported in part by the United States Department of Energy grant DE-SC0011637.

\bibliography{ref}

\end{document}